\documentclass[a4paper,11pt]{article}
\pdfoutput=1 

\usepackage{jheppub} 

\usepackage[T1]{fontenc}
\usepackage{multirow}

\allowdisplaybreaks
\definecolor{nicered}{rgb}{0.7,0.1,0.1}
\definecolor{nicegreen}{rgb}{0.1,0.5,0.1}
\definecolor{niceblue}{rgb}{0.0,0.1,0.7}
\hypersetup{colorlinks,citecolor=niceblue,linkcolor=niceblue,urlcolor=niceblue}

\usepackage[normalem]{ulem}

\def \bm#1{\mbox{\boldmath$#1$\unboldmath}}
\def \beq{\begin{equation}}
\def \eeq{\end{equation}}
\def \bea{\begin{eqnarray}}
\def \eea{\end{eqnarray}}

\title{Long-lived particle phenomenology \\ in the 2HDM+$\bm{a}$ model}

\author[a]{Ulrich Haisch}
\author[a, b]{and Luc Schnell}

\affiliation[a]{Max Planck Institute for Physics, \\ F{\"o}hringer Ring 6, 80805 M{\"u}nchen, Germany}
\affiliation[b]{Technische Universit{\"a}t M{\"u}nchen, Physik-Department, \\ James-Franck-Strasse 1, 85748 Garching, Germany}

\preprint{MPP-2023-27}

\emailAdd{haisch@mpp.mpg.de}
\emailAdd{schnell@mpp.mpg.de}

\abstract{Higgs decays displaced from the primary interaction vertex represent a striking experimental signature that is actively searched for by the ATLAS, CMS and LHCb collaborations. We point out that signals of this type appear in the context of the 2HDM+$a$~model if the mixing angle $\theta$ of the two CP-odd weak spin-0 eigenstates is tiny and the dark matter~(DM) sector is either decoupled or kinematically inaccessible. Utilising two suitable benchmark scenarios, we~determine the constraints on the parameter space of the 2HDM+$a$ model that are set by the existing LHC searches for long-lived particles~(LLPs) in Higgs decays. We find that depending on the precise mass spectrum of the spin-0 states, mixing angles $\theta$ in the ballpark of a few $10^{-8}$ to $10^{-5}$ can be excluded based on LHC~Run~II~data. This finding emphasises the unique role that searches for displaced signatures can play in constraining the parameter space of the 2HDM+$a$~model. The ensuing DM phenomenology is also discussed. In particular, we show that parameter choices leading to an interesting LLP phenomenology can simultaneously explain the DM abundance observed in today's~Universe.}

\begin{document} 
\maketitle
\flushbottom

\section{Introduction}
\label{sec:intro}

The two-Higgs-doublet plus pseudoscalar model~(2HDM+$a$)~\cite{Ipek:2014gua,No:2015xqa,Goncalves:2016iyg,Bauer:2017ota} has by now established itself as a pillar of the LHC dark matter~(DM) search programme~\cite{ATLAS:2017hoo,Pani:2017qyd,Tunney:2017yfp,Arcadi:2017wqi,LHCDarkMatterWorkingGroup:2018ufk,CMS:2018zjv,ATLAS:2019wdu,Arcadi:2020gge,CMS:2020ulv,ATLAS:2020yzc,ATLAS:2021jbf,ATLAS:2021shl,ATLAS:2021gcn,ATLAS:2021fjm,Robens:2021lov,Argyropoulos:2021sav,Banerjee:2021hfo,Argyropoulos:2022ezr,ATLAS:2022ecu,ATLAS:2022znu,CMS:2022sfl,Arcadi:2022lpp}. It includes a DM candidate in the form of a Dirac fermion which is a singlet under the Standard Model~(SM) gauge group, four 2HDM spin-0 states and an additional CP-odd mediator that is meant to provide the dominant portal between the dark and the visible sector. Since in models with pseudoscalar mediators the DM~direct detection constraints are weaker compared to models with scalar mediators, the former models are more attractive from an astrophysical point of view since they often allow to reproduce the observed DM relic abundance in a wider parameter space and with less tuning. These features admit a host of missing transverse momentum~($E_T^{\rm miss}$) and non-$E_T^{\rm miss}$ signatures in the 2HDM+$a$ model at colliders which can and have been consistently compared and combined. See for instance~\cite{Argyropoulos:2021sav,ATLAS:2022rxn,Argyropoulos:2022ezr} for such combinations. 

Beyond the SM~(BSM) scenarios in which the hidden and the visible sectors are connected through a Higgs portal are also being actively probed for at colliders. One rather generic feature in such BSM models is the appearance of new electrically neutral long-lived particles~(LLPs) that give rise to displaced vertex signatures in the LHC detectors --- see~for~example~\cite{Curtin:2018mvb,Lee:2018pag,Alimena:2019zri} for detailed reviews of theoretical and experimental aspects of~LLPs at the LHC. The main goal of this article is to point out that besides interesting prompt $E_{T, \rm miss}$ and non-$E_{T, \rm miss}$ signatures, the 2HDM+$a$ model can also have an attractive~LLP~phenomenology. In fact, in the 2HDM+$a$ model the role of the LLP is played by the additional pseudoscalar $a$, which depending on its mass can be pair produced efficiently in the decays of both the $125 \, {\rm GeV}$ Higgs and the non-SM CP-even Higgs,~i.e.~$h \to aa$ and $H \to aa$. To~illustrate the different facets of the LLP phenomenology in the 2HDM+$a$ model we identify two suitable parameter benchmarks. For these benchmark scenarios we determine the bounds on the mixing angle $\theta$ of the two CP-odd weak spin-0 eigenstates as a function of the LLP mass that are set by the existing LHC searches for displaced Higgs decays~\cite{LHCb:2017xxn,ATLAS:2018pvw,ATLAS:2018tup,ATLAS:2019qrr,ATLAS:2019jcm,ATLAS:2020ahi,CMS:2020iwv,CMS:2021juv,ATLAS:2021jig,CMS:2021kdm,CMS:2021yhb,CMS:2021sch,ATLAS:2022gbw,ATLAS:2022zhj,CMS:2022qej,ATL-PHYS-PUB-2022-007,CMSLLP}. It~turns out that depending on the precise mass spectrum of the spin-0 states, mixing angles~$\theta$ from around a few $10^{-8}$ to about $10^{-5}$ can be excluded with LHC~Run~II data. To the best of our knowledge, mixing angles $\theta$ in this range cannot be tested by any other means, which highlights the special role that LLPs searches play in constraining the parameter space of the 2HDM+$a$~model. In fact, as we will further demonstrate, parameter choices that lead to an interesting LLP phenomenology can in general also correctly predict the measured DM relic density. The regions of 2HDM+$a$ parameter space singled out in our article therefore deserve, in our humble opinion, dedicated experimental explorations in future LHC runs. 

This work is structured as follows: in Section~\ref{sec:prelim} we detail the theoretical ingredients that are relevant in the context of this article. Our general findings concerning the LLP phenomenology in the 2HDM+$a$ model will be illustrated in Section~\ref{sec:pheno} by considering two suitable parameter benchmark scenarios as examples. For these two benchmark choices we derive in Section~\ref{sec:bounds} the constraints that the existing LHC searches for LLPs in displaced Higgs decays place on the 2HDM+$a$ parameter space. In Section ~\ref{sec:relic} we discuss the resulting DM phenomenology. Section~\ref{sec:summary}~concludes our work. 

\section{2HDM+$\bm{a}$ model primer}
\label{sec:prelim}

In order to understand under which circumstances the additional pseudoscalar $a$ in the 2HDM+$a$ model can be an~LLP it is useful to recall its partial decay modes --- further details on the structure of the 2HDM+$a$ model can be found for instance in~\cite{Bauer:2017ota,LHCDarkMatterWorkingGroup:2018ufk}. In the alignment limit,~i.e.~$\cos \left (\beta - \alpha \right) = 0$, and choosing for concreteness the Yukawa sector of the 2HDM+$a$~model to be of type-II, one has at tree level 
\beq \label{eq:Gammaa}
\begin{split}
\Gamma \left ( a \to \chi \bar \chi \right ) & = \frac{y_\chi^2}{8 \hspace{0.125mm} \pi} \, m_a\, \sqrt{ 1 - \frac{4 \hspace{0.125mm} m_\chi^2}{m_a^2}} \, \cos^2 \theta \,, \\[2mm]
\Gamma \, ( a \to f \bar f ) & = \frac{N_c^f \eta_f^2 \hspace{0.5mm} y_f^2}{16\hspace{0.125mm} \pi} \, m_a\, \sqrt{ 1 - \frac{ 4 \hspace{0.125mm} m_f^2}{m_a^2}} \, \sin^2 \theta \,.
\end{split}
\eeq
At the one-loop level the pseudoscalar $a$ can also decay to gauge bosons. The largest partial decay width is the one to gluon pairs. It takes the form
\beq \label{eq:Gammag}
\Gamma \left ( a \to g g\right ) = \frac{\alpha_s^2}{32\hspace{0.125mm} \pi^3 \hspace{0.125mm} v^2} \, m_a^3 \, \left | \sum_{q=t,b,c} \eta_q \, f \left ( \frac{4 \hspace{0.125mm} m_q^2}{m_a^2} \right ) \right |^2 \, \sin^2 \theta \,,
\eeq
with 
\beq \label{eq:fx}
f(z) = z \arctan^2 \left ( \frac{1}{\sqrt{z -1}} \right ) \,.
\eeq
Here $m_a$ is the mass of the pseudoscalar $a$, $m_\chi$ is the mass of the DM particle, $y_\chi$ is the Yukawa coupling of the pseudoscalar $a$ to a pair of DM particles and $\sin \theta$ quantifies the mixing of the two CP-odd weak spin-0 eigenstates. Furthermore, $N_c^q = 3$, $N_c^l = 1$, $\eta_{u} = \cot \beta$, $\eta_{d} = \tan \beta$, $\eta_{l} = \tan \beta$ and $y_f = \sqrt{2} \hspace{0.25mm} m_f/v$ with $m_f$ the mass of the relevant SM fermion, $v \simeq 246 \, {\rm GeV}$ the Higgs vacuum expectation value~(VEV) and $\alpha_s$ the strong coupling constant. From the analytic expressions~(\ref{eq:Gammaa}) and~(\ref{eq:Gammag}) it is evident that the pseudoscalar~$a$ can only be long-lived if $\sin \theta$ is sufficiently small, i.e.~$\sin \theta \to 0$, and decays to DM are strongly suppressed/absent which can be achieved either via decoupling,~i.e.~$y_\chi \to 0$, or by forbidding the process kinematically,~i.e.~$m_\chi > m_a/2$. 

Given the strong suppression of the couplings of the pseudoscalar $a$ to SM fermions in the limit $\sin \theta \to 0$, the only possibility to produce a long-lived~$a$ is via the decay of heavier spin-0 state~$\phi$ into a pair of such pseudoscalars. In the case that the scalar potential is CP~conserving the~$\phi$ has to be a CP-even state which implies that in the 2HDM+$a$ model one can have both decays of the $125 \, {\rm GeV}$ Higgs $h$ and the heavy CP-even Higgs $H$. The~corresponding partial decay widths can be written as 
\beq \label{eq:Gammaphi}
\Gamma \left ( \phi \to a a \right ) = \frac{g_{\phi aa}^2}{32 \hspace{0.125mm} \pi} \, m_\phi \, \sqrt{ 1 - \frac{ 4 \hspace{0.125mm} m_a^2}{m_\phi^2}} \,, 
\eeq
with $\phi = h,H$. For $\sin \theta \simeq 0$ the relevant trilinear couplings are given by~\cite{Bauer:2017ota}
\beq \label{eq:ghHaa}
\begin{split}
g_{haa} & \simeq -\frac{2 \hspace{0.25mm} v}{m_h} \left (\lambda_{P1} \cos^2 \beta + \lambda_{P2} \sin^2 \beta \right ) \,, \\[2mm]
g_{Haa} & \simeq \frac{v}{m_H} \sin \left ( 2 \beta \right ) \left (\lambda_{P1} - \lambda_{P2} \right ) \,,
\end{split}
\eeq
where $m_h \simeq 125 \, {\rm GeV}$ is the mass of the SM-like Higgs, while $\lambda_{P1}$ and $\lambda_{P2}$ are the quartic couplings that appear in the 2HDM+$a$ scalar potential as follows $P^2 \hspace{0.25mm} \big ( \lambda_{P1} \hspace{0.25mm} H_1^\dagger H_1 + \lambda_{P2} \hspace{0.25mm} H_2^\dagger H_2 \big )$ (see for example~\cite{Bauer:2017ota,LHCDarkMatterWorkingGroup:2018ufk} for the complete expression of the scalar potential). Here $P$ denotes the additional pseudoscalar in the weak eigenstate~basis which satisfies $P\simeq a$ for $\sin \theta \simeq 0$. 

\begin{figure}[t!]
\begin{center}
\includegraphics[width=0.75\textwidth]{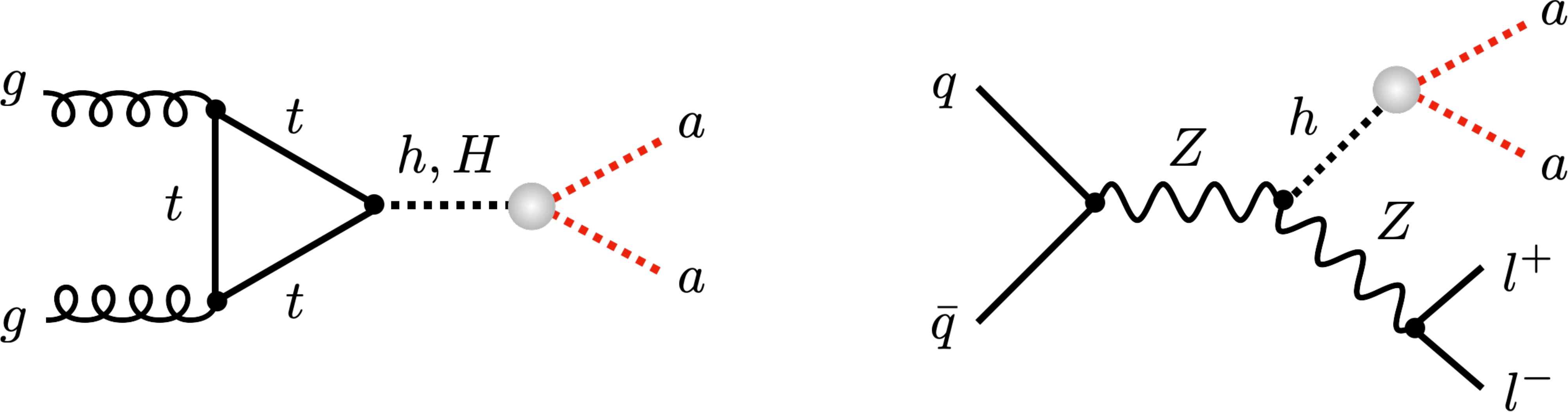}
\end{center}
\vspace{0mm} 
\caption{\label{fig:diagrams1} Examples of tree-level Feynman diagrams representing $pp \to aa$ production via gluon-gluon-fusion~(ggF) Higgs production~(left) and $pp \to l^+ l^- a a$ production in associated $Zh$ production~(right) in the 2HDM+$a$~model. The possible decay modes of the pseudoscalar $a$ are not shown. Consult~the text for further details. }
\end{figure}
 
The trilinear couplings~entering~(\ref{eq:ghHaa}) can be constrained phenomenologically. In the case of $g_{haa}$ one can require that the partial decay width $\Gamma \left (h \to aa \right)$ does not exceed the total decay width $\Gamma_h$ of the $125 \, {\rm GeV}$ Higgs as measured directly at the LHC. For $m_a \ll m_h$ this leads to the inequality~\cite{Argyropoulos:2022ezr} 
\beq \label{eq:boundghaa}
\left | g_{haa} \right | \lesssim \sqrt{\frac{32 \hspace{0.25mm} \pi \hspace{0.25mm} \Gamma_h }{m_h}} \simeq 0.94 \,,
\eeq
where in the last step we have employed the latest 95\% confidence level~(CL) bound of $\Gamma_h < 1.1 \, {\rm GeV}$ that follows from the direct LHC measurements of the total SM-like Higgs width~\cite{CMS:2017dib,ATLAS:2018tdk}. Inserting the first expression of~(\ref{eq:ghHaa}) into~(\ref{eq:boundghaa}) then leads to the following relation
\beq \label{eq:bound1}
\left | \lambda_{P1} \cos^2 \beta + \lambda_{P2} \sin^2 \beta \right | \lesssim 0.24 \,.
\eeq

In the case of $g_{Haa}$ one obtains in a similar fashion 
\beq \label{eq:boundgHaa}
\left | g_{Haa} \right | \lesssim \sqrt{\frac{32 \hspace{0.25mm} \pi \hspace{0.25mm} \Gamma_H }{m_H}} \simeq 3.2 \,,
\eeq
where in the final step we have set the ratio between the total decay width and the mass of the heavy Higgs to $\Gamma_H/m_H = 10\%$. This choice is motivated by the observation that for significantly larger ratios different treatments of the $H$ propagator lead to notable changes in the heavy Higgs production cross section and thus the~LLP signal compared to the case of a Breit-Wigner propagator with fixed width. Combining the second relation in~(\ref{eq:ghHaa}) with~(\ref{eq:boundgHaa}) it then follows that 
\beq \label{eq:bound2}
\left | \frac{v}{m_H} \sin \left ( 2 \beta \right ) \left (\lambda_{P1} - \lambda_{P2} \right ) \right | \lesssim 3.2 \,.
\eeq

The above discussion should have shown that in the limit $\sin \theta \to 0$, an~LLP signature can arise in the 2HDM+$a$ model from~$h$~or~$H$ production followed by the decay of the intermediate Higgs to a pair of pseudoscalars. Representative tree-level graphs showing $pp \to aa$~production in ggF Higgs production~(left) and $pp \to l^+ l^- a a$ production in associated Higgs production~(right) that appear in the 2HDM+$a$ model can be found in~Figure~\ref{fig:diagrams1}. Notice that in the former case both the $125 \, {\rm GeV}$ Higgs and the heavy CP-even Higgs contribute in the alignment limit. This is not the case for the latter process as the~$HZZ$ vertex vanishes identically in the limit $\cos \left ( \beta - \alpha \right) \to 0$. 

\section{Parameter benchmarks}
\label{sec:pheno}

Besides the phenomenological bounds~(\ref{eq:bound1}) and (\ref{eq:bound2}) that constrain the sizes of the couplings $\lambda_{P1}$ and $\lambda_{P2}$, the requirement for the scalar potential to be bounded from below also restricts the quartic couplings as well as other parameters of the 2HDM+$a$~model. Assuming that $\lambda_{P1}, \lambda_{P2} >0$ and that $\sin \theta \simeq 0$, one finds two bounded from below conditions that take the form~\cite{LHCDarkMatterWorkingGroup:2018ufk}
\beq \label{eq:BFB}
\lambda_3 > 2 \lambda \,, \qquad 
\lambda_3 \gtrsim -2 \lambda \cot^2 \left ( 2 \beta \right ) \,.
\eeq
Here the parameter $\lambda_3$ denotes the usual quartic coupling from the 2HDM scalar potential and $\lambda = m_h^2/(2 \hspace{0.25mm} v^2) \simeq 0.13$ is the cubic SM Higgs self-coupling. In order to fulfill these relations and to avoid the tight constraints from Higgs and electroweak precision physics, we make the following common parameter choices 
\beq \label{eq:basicchoices}
\lambda_3 = 0.3 \,, \quad \cos \left ( \beta - \alpha \right ) = 0 \,, \quad \tan \beta = 1 \,, \quad m_H = m_A = m_{H^\pm} \,, \quad y_\chi = 1\,.
\eeq 
We furthermore employ a Yukawa sector of type-II throughout this work.

The first 2HDM+$a$ benchmark scenario that we will study as an example to illustrate the possible LLP phenomenology in the 2HDM+$a$ model is:
\beq \label{eq:benchmarkI}
\big \{ \lambda_{P1}, \lambda_{P2} ,m_\chi \big\} = \big \{ 2 \cdot 10^{-3}, 2 \cdot 10^{-3} , 170 \, {\rm GeV} \big \} \,, \quad (\text{benchmark~I}) \,.
\eeq
We furthermore treat $\sin \theta$ and $m_a$ as free parameters but require that $m_a < m_h/2$ so that the LLP can be pair produced in the decay of the $125 \, {\rm GeV}$ Higgs boson. The precise value of the common heavy Higgs mass is irrelevant in such a situation and we simply set it to $m_H = 600 \, {\rm GeV}$ in benchmark~I. Notice that the quartic couplings $\lambda_{P1}$ and $\lambda_{P2}$ have been chosen such that the constraint~(\ref{eq:bound1}) is easily fulfilled. In fact, in the limit $m_a \to 0$ the benchmark~I parameter choices imply 
\beq \label{eq:Gammah}
\Gamma_h = 4.15 \, {\rm MeV} \,, 
\eeq
a value that is very close to the SM prediction of $\Gamma_h^{\rm SM} = 4.07 \, {\rm MeV}$~\cite{ParticleDataGroup:2020ssz}. The corresponding $h \to aa$ branching ratio is 
\beq \label{eq:BRhaa}
{\rm BR} \left ( h \to aa \right ) = 1.9\% \,.
\eeq
The proper decay length of the pseudoscalar $a$ for masses in the range $m_a \in [20, 60 ] \, {\rm GeV}$ can be approximated by 
\beq \label{eq:ctI}
\frac{c \hspace{0.125mm} \tau_a}{\rm m} \simeq 4.8 \cdot 10^{-12} \, \left ( \frac{\rm GeV}{m_a} \right )^{0.9} \, \frac{1}{\sin^2 \theta} \,, 
\eeq
which means that for 
\beq \label{eq:sinthetaI}
\sin \theta \simeq 4.2 \cdot 10^{-7} \,, 
\eeq
a pseudoscalar of $m_a = 40 \, {\rm GeV}$ has a proper decay length of around $1 \hspace{0.25mm} {\rm m}$. The result~(\ref{eq:ctI}) includes higher-order QCD corrections employing the formulas presented in Appendix~A of the paper~\cite{Haisch:2018kqx} as implemented in~\cite{2HDMawidthcalc}. Also notice that for the choices~(\ref{eq:benchmarkI}) and assuming that $\sin \theta \simeq 0$, the additional 2HDM Higgses are all narrow,~i.e.~$\Gamma_H/m_H \simeq 2\%$, $\Gamma_A/m_A \simeq 4\%$ and $\Gamma_{H^\pm}/m_{H^\pm} \simeq 4\%$, with ${\rm BR} \left (H \to t \bar t \right ) \simeq 100\%$, ${\rm BR} \left (A \to t \bar t \right ) \simeq 100\%$ and ${\rm BR} \left (H^\pm \to t b \right ) \simeq 100\%$. 

In our second 2HDM+$a$ benchmark scenario that leads to an interesting LLP phenomenology, we consider the following parameter choices
\beq \label{eq:benchmarkII}
\big \{ \lambda_{P1}, \lambda_{P2}, m_\chi \big \} = \big \{ 3, 0, 770 \, {\rm GeV} \big \} \,, \quad (\text{benchmark~II}) \,.
\eeq
The parameters $\sin \theta$, $m_H$ and $m_a$ are instead treated as input with the requirements that $m_a > m_h/2$ and $m_a < m_H/2$ so that the LLP can only be pair produced in the decay $H \to aa$ of the heavy CP-even Higgs boson $H$. Notice that the values $\lambda_{P1}$ and $\lambda_{P2}$ in~(\ref{eq:benchmarkII}) satisfy the constraint~(\ref{eq:bound2}). Taking for example $m_H = 600 \, {\rm GeV}$ and $m_a = 150 \, {\rm GeV}$, the total decay width of the heavy CP-even Higgs is given by 
\beq \label{eq:Gammah}
\Gamma_H = 22 \, {\rm GeV} \,, 
\eeq
which implies that $\Gamma_H/m_H = 3.7\%$. The corresponding branching ratios are 
\beq \label{eq:BRHaa}
{\rm BR} \left ( H \to aa \right ) = 35\% \,, \qquad {\rm BR} \left ( H \to t \bar t \right ) = 65\% \,, 
\eeq
meaning that the decays of the heavy Higgs to two LLPs does not have the largest branching ratio but that di-top decays are more frequent. Notice that given the structure of the trilinear coupling~$g_{Haa}$ in (\ref{eq:ghHaa}) this feature will be even more pronounced for heavier CP-even Higgs bosons $H$. In the range $m_a \in [100, 300 ] \, {\rm GeV}$, the proper decay length of the pseudoscalar~$a$ is approximately given by 
\beq \label{eq:ctII}
\frac{c \hspace{0.125mm} \tau_a}{\rm m} \simeq 1.2 \cdot 10^{-10} \, \left ( \frac{\rm GeV}{m_a} \right )^{1.6} \, \frac{1}{\sin^2 \theta} \,,
\eeq
where again the results of~\cite{Haisch:2018kqx,2HDMawidthcalc} have been used. It follows that for 
\beq \label{eq:sinthetaI}
\sin \theta \simeq 2.1 \cdot 10^{-7} \,, 
\eeq
a pseudoscalar of $m_a = 150 \, {\rm GeV}$ has a proper decay length of about $1 \hspace{0.25mm} {\rm m}$. Notice finally that in the case of~(\ref{eq:benchmarkI}) with $\sin \theta \simeq 0$, one has $\Gamma_A/m_A \simeq 4\%$ and $\Gamma_{H^\pm}/m_{H^\pm} \simeq 4\%$ with ${\rm BR} \left (A \to t \bar t \right ) \simeq 100\%$ and ${\rm BR} \left (H^\pm \to t b \right ) \simeq 100\%$. 

\section{LLP constraints}
\label{sec:bounds}

At the LHC, searches for displaced Higgs boson decays into LLPs have been carried out by the ATLAS, CMS and LHCb collaborations in different final states, covering proper decay lengths from around $10^{-3} \hspace{0.5mm} {\rm m}$ to $10^{3} \hspace{0.5mm} {\rm m}$~\cite{LHCb:2017xxn,ATLAS:2018pvw,ATLAS:2018tup,ATLAS:2019qrr,ATLAS:2019jcm,ATLAS:2020ahi,CMS:2020iwv,CMS:2021juv,ATLAS:2021jig,CMS:2021kdm,CMS:2021yhb,CMS:2021sch,ATLAS:2022gbw,ATLAS:2022zhj,CMS:2022qej,ATL-PHYS-PUB-2022-007,CMSLLP}. The LLP mean decay length determines the search strategies and reconstruction techniques that are employed --- see for instance~Section~5~of the review~\cite{Argyropoulos:2021sav} for comprehensive descriptions of the details of the experimental techniques employed in LHC LLP searches. 

We first consider the 2HDM+$a$ benchmark~I~scenario~(\ref{eq:benchmarkI}) with $m_a < m_h/2$. In this case the pseudoscalar $a$ can be pair produced in the decay of the $125 \, {\rm GeV}$ Higgs boson. Depending on whether the mass of the $a$ is below or above the bottom-quark threshold, the dominant decay modes of the pseudoscalar are ${\rm BR} \left ( a \to c \bar c \right ) \simeq 53\%$, ${\rm BR} \left ( a \to \tau^+ \tau^- \right ) \simeq 38\%$ and ${\rm BR} \left ( a \to g g \right ) \simeq 10\%$ or ${\rm BR} \, ( a \to b \bar b ) \simeq 85\%$, ${\rm BR} \left ( a \to c \bar c \right ) \simeq 4\%$, ${\rm BR} \left ( a \to \tau^+ \tau^- \right ) \simeq 7\%$ and ${\rm BR} \left ( a \to g g \right ) \simeq 3\%$, respectively. LLP~searches that target pseudoscalar pair production in ggF Higgs or associated~$Zh$~production (cf.~Figure~\ref{fig:diagrams1}) leading to multi-jet or four-bottom final states therefore provide the most stringent constraints. Looking for displaced leptons instead leads to significantly weaker restrictions because of the small leptonic branching ratios. 

\begin{figure}[t!]
\begin{center}
\includegraphics[width=0.6\textwidth]{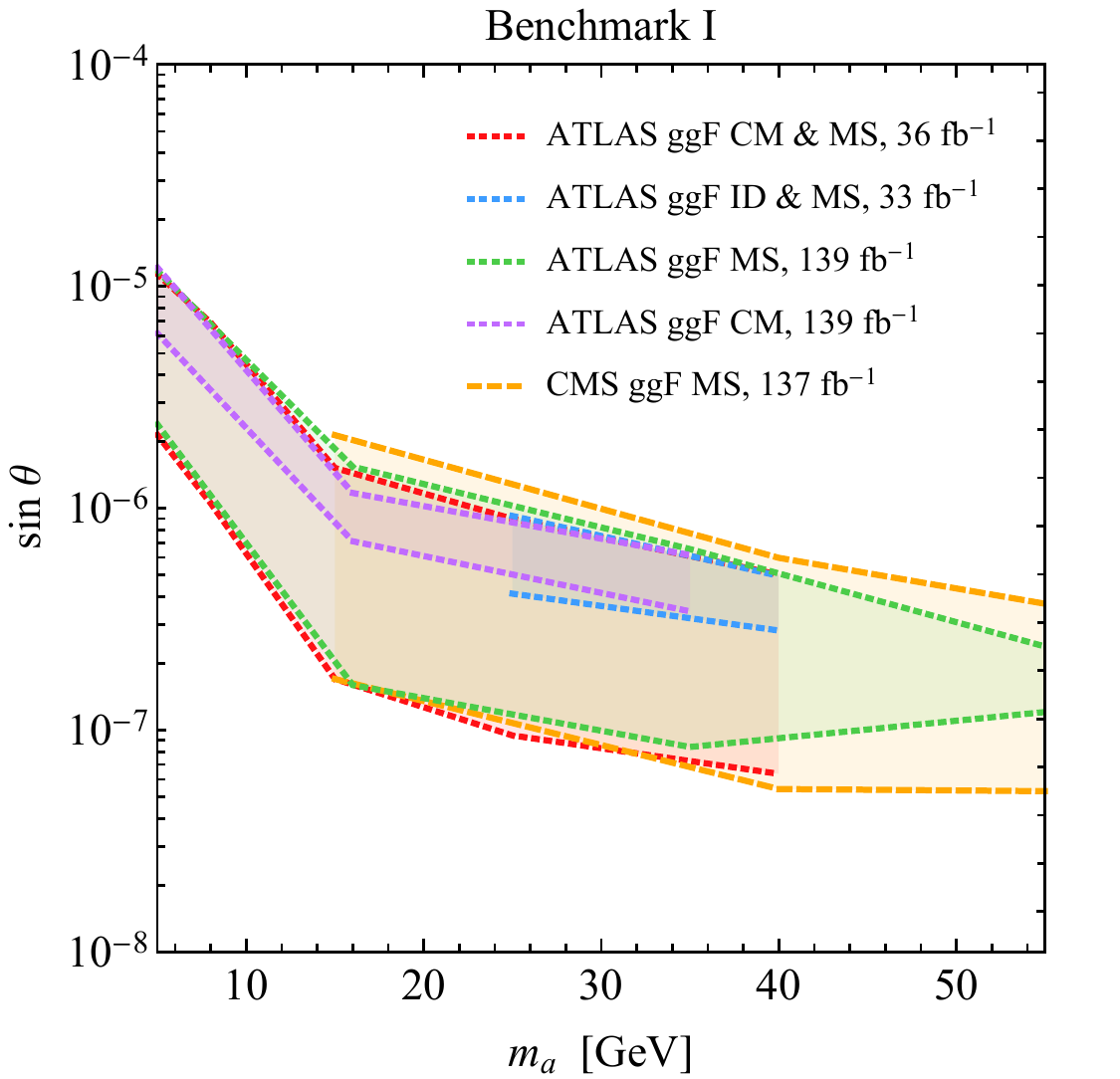}
\end{center}
\vspace{-4mm} 
\caption{\label{fig:summary1} 95\%~CL exclusion regions in the $m_a\hspace{0.25mm}$--$\hspace{0.5mm} \sin \theta$ plane for the 2HDM+$a$ benchmark~I scenario~(\ref{eq:benchmarkI}). The~dotted red, blue, green and purple lines correspond to the limits following from the ATLAS searches~\cite{ATLAS:2018tup,ATLAS:2019qrr}, \cite{ATLAS:2019jcm}, \cite{ATLAS:2022gbw} and \cite{ATLAS:2022zhj}, respectively. The dashed yellow curves instead represent the bound that arises from the CMS search~\cite{CMS:2021juv}. The parameter space between the lines is disfavoured. See~main text for further details.}
\end{figure}

In~Figure~\ref{fig:summary1} we show an assortment of LLP constraints in the $m_a\hspace{0.25mm}$--$\hspace{0.5mm} \sin \theta$ plane that apply in the case of~(\ref{eq:benchmarkI}). All limits result from LHC searches that consider ggF Higgs production. The~dotted red exclusion corresponds to the search~\cite{ATLAS:2019qrr} that considers displaced hadronic jets in the ATLAS calorimeter~(CM) and the muon spectrometer~(MS)~\cite{ATLAS:2018tup}, while the dotted blue constraint instead results from the ATLAS search~\cite{ATLAS:2019jcm} that utilises the inner detector~(ID) and the MS. These searches use up to $36 \, {\rm fb}^{-1}$ and $33 \, {\rm fb}^{-1}$ of $\sqrt{s} = 13 \, {\rm TeV}$ data, respectively. The dotted green (purple) lines represent an upgrade of the MS (CM) search strategy to $139 \, {\rm fb}^{-1}$ of luminosity collected in LHC~Run~II. The~corresponding limits are reported in the ATLAS publication~\cite{ATLAS:2022gbw} and~\cite{ATLAS:2022zhj}, respectively. The dashed yellow contour is finally the exclusion that derives from the CMS search~\cite{CMS:2021juv} which employs the muon endcap and $137 \, {\rm fb}^{-1}$ of $\sqrt{s} = 13 \, {\rm TeV}$ data. From the~figure it is evident that in the 2HDM+$a$ benchmark~I~scenario the existing LHC searches for displaced Higgs decays to hadronic jets allow to exclude values of $\sin \theta$ between around $10^{-7}$ and $10^{-5}$ with the exact bound depending on the mass of the pseudoscalar $a$. The excluded parameter space corresponds to proper decay lengths~$c \hspace{0.125mm} \tau_a$ in the range from around $59 \hspace{0.5mm}{\rm m}$ to $0.08 \hspace{0.25mm} {\rm m}$. Notice that given the smallness of the $h \to aa$ branching ratio $\big($cf.~(\ref{eq:BRhaa})$\big)$, our benchmark~I scenario easily evades the present bounds on the undetected or invisible branching ratios of the $125 \, {\rm GeV}$ Higgs~\cite{ATLAS-CONF-2020-027} that amount to 19\% and 9\%, respectively. In fact, even a possible future high-luminosity LHC~(HL-LHC) upper limit on the invisible branching ratio of the SM-like Higgs of ${\rm BR} \left ( h \to {\rm invisible} \right) < 2.5\%$~\cite{Cepeda:2019klc} would not be stringent enough to test~(\ref{eq:benchmarkI}) indirectly. This feature underlines the special role that LLP searches for displaced Higgs decays can play in testing 2HDM+$a$ models with mixing angles $\theta$ close to zero. 

\begin{figure}[t!]
\begin{center}
\includegraphics[width=0.6\textwidth]{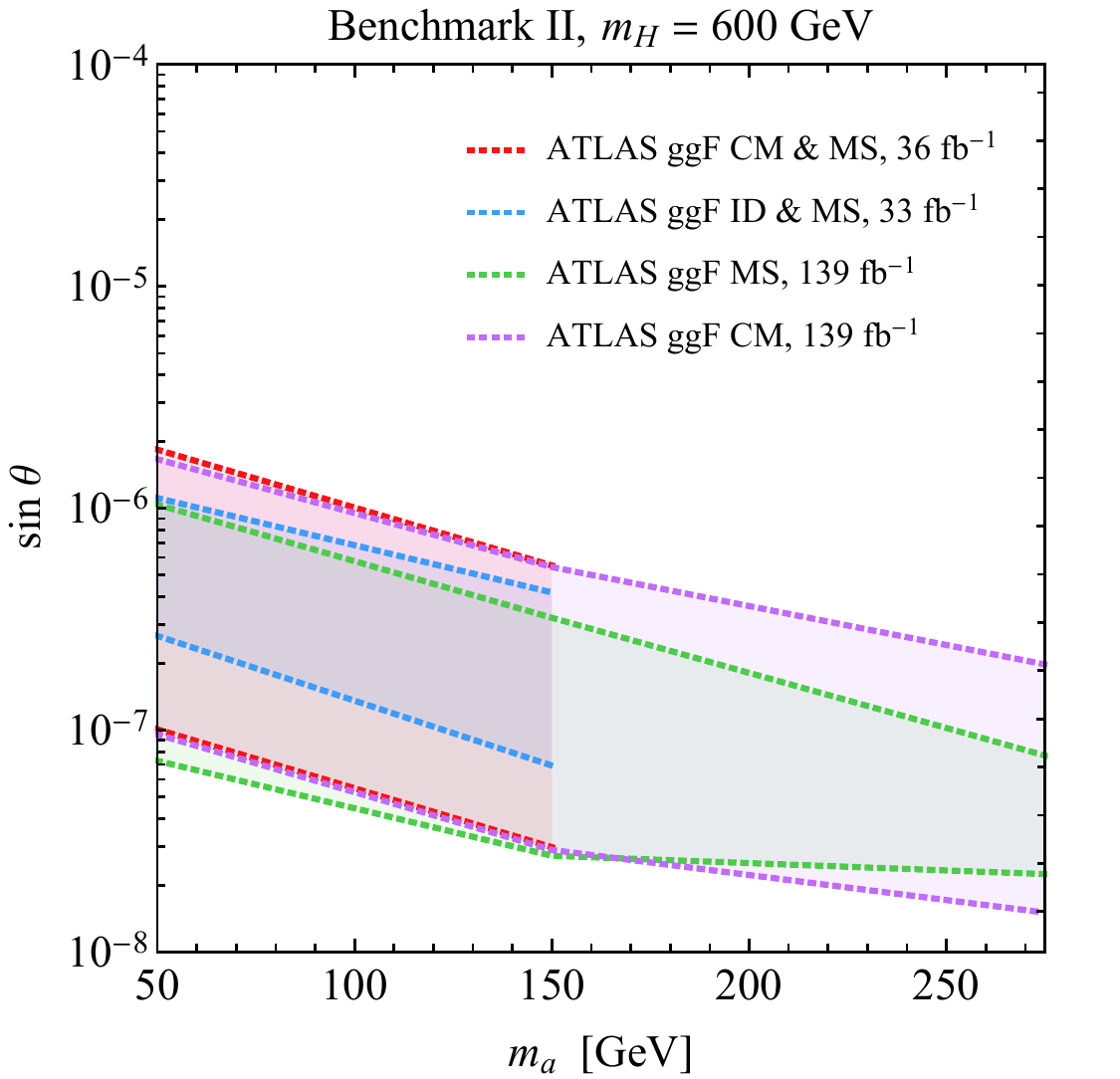}

\hspace{6mm}

\includegraphics[width=0.6\textwidth]{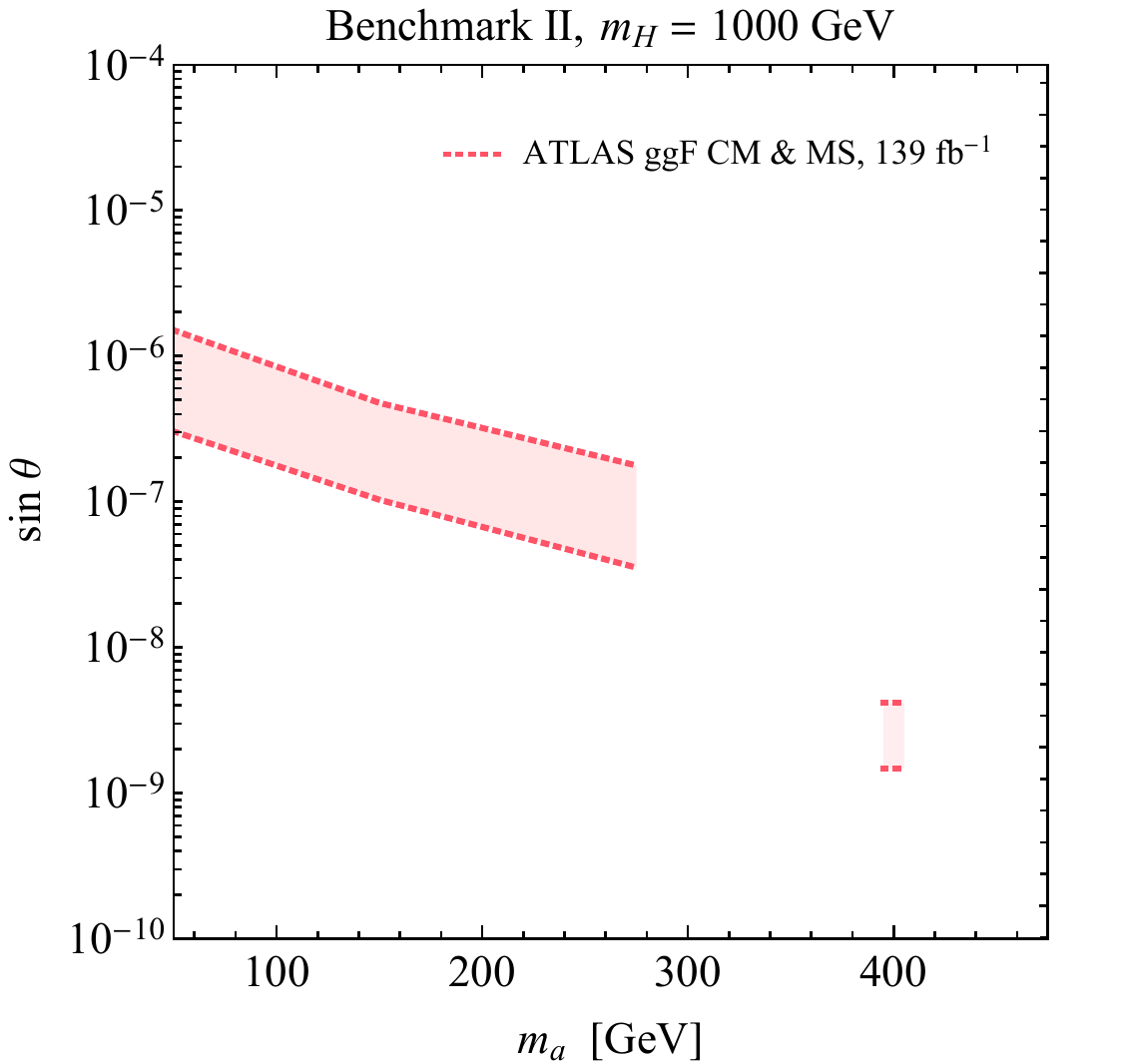}
\end{center}
\vspace{-4mm} 
\caption{\label{fig:summary23} As Figure~\ref{fig:summary1} but for the two 2HDM+$a$ benchmark~II scenarios~(\ref{eq:benchmarkII}). The upper~(lower) panel depicts the results for $m_H = 600 \, {\rm GeV}$ ($m_H = 1000 \, {\rm GeV}$). The~dotted red, blue, green and purple lines in the upper panel correspond to the bounds following from~\cite{ATLAS:2018tup,ATLAS:2019qrr}, \cite{ATLAS:2019jcm}, \cite{ATLAS:2022gbw} and~\cite{ATLAS:2022zhj}, respectively. The dotted red exclusion in the lower panel instead represents the combination of the ATLAS searches~\cite{ATLAS:2018tup,ATLAS:2019qrr,ATLAS:2022gbw,ATLAS:2022zhj}. The shaded parameter regions are disfavoured. Further explanations can be found in the main text.}
\end{figure}

Let us now turn our attention to the benchmark~II~scenario~(\ref{eq:benchmarkII}). In this case the parameters are chosen such that an LLP signal may arise from the prompt decay of the heavy CP-even Higgs,~i.e~$H \to aa$, followed by the displaced decays of the pseudoscalars to a pair of SM fermions $a \to f \bar f$ or gluons $a \to gg$. Given our choice of Yukawa sector and $\tan \beta$, the $a$ dominantly decays to the heaviest SM fermion, which means that depending on the precise value of its mass either $a \to b \bar b$ or $a \to t \bar t$ provide the largest rate. To illustrate these two possibilities we consider in benchmark~II the mass combination $m_H = 600 \, {\rm GeV}$ with $m_a \in [50, 275] \, {\rm GeV}$ as well as $m_H = 1000 \, {\rm GeV}$ with $m_a \in[50, 475] \, {\rm GeV}$. At a centre-of-mass energy of $\sqrt{s} = 13 \,{\rm TeV}$ the relevant inclusive heavy Higgs production cross sections are $\sigma \left ( p p \to H \right ) \simeq 2.0 \, {\rm pb}$ and $\sigma \left ( p p \to H \right ) \simeq 0.12 \, {\rm pb}$~\cite{HiggsWG}, respectively. Notice that in the first case and assuming $m_a = 150 \, {\rm GeV}$, the heavy Higgs branching ratios~(\ref{eq:BRHaa}) as well as ${\rm BR} \, ( a \to b \bar b ) \simeq 62\%$, ${\rm BR} \left ( a \to c \bar c \right ) \simeq 3\%$, ${\rm BR} \left ( a \to \tau^+ \tau^- \right ) \simeq 7\%$ and ${\rm BR} \left ( a \to g g \right ) \simeq 28\%$ apply. In the second case, one has instead ${\rm BR} \left ( H \to a a \right ) \simeq 9\%$, ${\rm BR} \left ( H \to t \bar t \right ) \simeq 91\%$ and ${\rm BR} \left ( a \to t \bar t \right ) \simeq 100\%$ for $m_a = 400 \, {\rm GeV}$. 

In the upper panel of~Figure~\ref{fig:summary23} we display the relevant 95\%~CL exclusion regions in the $m_a\hspace{0.25mm}$--$\hspace{0.5mm} \sin \theta$ plane that apply in the case of the 2HDM+$a$ benchmark~II scenario with $m_H = 600 \, {\rm GeV}$. One observes that taken together the ATLAS searches~\cite{ATLAS:2018tup,ATLAS:2019qrr,ATLAS:2019jcm,ATLAS:2022gbw,ATLAS:2022zhj} allow to exclude $\sin \theta$ values between $2 \cdot 10^{-8}$ and $2 \cdot 10^{-6}$. The corresponding proper decay lengths~$c \hspace{0.125mm} \tau_a$ range from $53 \hspace{0.5mm}{\rm m}$ to $0.04 \hspace{0.25mm} {\rm m}$. The lower plot in~Figure~\ref{fig:summary23} shows the limits on $\sin \theta$ that a combination of the four ATLAS searches~\cite{ATLAS:2018tup,ATLAS:2019qrr,ATLAS:2022gbw,ATLAS:2022zhj} allow to set in the 2HDM+$a$ benchmark~II scenario~(\ref{eq:benchmarkII}) assuming $m_H = 1000 \, {\rm GeV}$. One observes that the existing LLP searches can exclude mixing angles for the mass points $m_a = 50 \, {\rm GeV}, 150 \, {\rm GeV}, 275 \, {\rm GeV}$ and $400 \, {\rm GeV}$, while the 2HDM+$a$ realisation with $m_a = 475 \, {\rm GeV}$ remains untested at present. For~pseudoscalar masses below the top-quark threshold, $\sin \theta$ values between around~$4 \cdot 10^{-8}$ and~$2 \cdot 10^{-6}$ are excluded, whereas for $m_a = 400 \, {\rm GeV}$ mixing parameters in the range of about~$4 \cdot 10^{-9}$ and~$1 \cdot 10^{-9}$ are disfavoured. The excluded parameter space corresponds to $c \hspace{0.125mm} \tau_a$ values ranging from around $9.7 \hspace{0.25mm}{\rm m}$ to $0.06 \hspace{0.25mm} {\rm m}$. Notice that the order of magnitude improvement of the constraint on $\sin \theta$ from the point $m_a = 275 \, {\rm GeV}$ to $m_a = 400 \, {\rm GeV}$ is readily understood by recalling that in the former case one has $\Gamma_a \simeq 16 \, {\rm MeV} \sin^2 \theta$, while in the latter case $\Gamma_a \simeq 12 \, {\rm GeV} \sin^2 \theta$ as a result of the open $a \to t \bar t$ channel. We add that improving the limits \cite{ATLAS:2022gbw,ATLAS:2022zhj} by a factor of four would allow to probe our 2HDM+$a$ benchmark~II scenario for $m_H = 1000 \, {\rm GeV}$ and $m_a = 475 \, {\rm GeV}$. A final remark concerns the possibility to search for the heavy CP-even Higgs in the processes $pp \to H \to t \bar t$ or $pp \to t \bar t H \to 4t$. While LHC searches for spin-0 resonances in both di-top~\cite{ATLAS:2017snw,ATLAS:2018alq,CMS:2019pzc} and four-top production~\cite{CMS:2019rvj,ATLAS:2020hpj,ATLAS:2021kqb,CMS-PAS-TOP-21-005} have been performed, it turns out that the existing searches do not provide any bound on our 2HDM+$a$ benchmark~II model for both $m_H = 600 \, {\rm GeV}$ and $m_H = 1000 \, {\rm GeV}$ in the $m_a$ ranges considered in~Figure~\ref{fig:summary23}. This finding again stresses the unique role that LLPs searches can play in the 2HDM+$a$ model in constraining the parameter space. 

\section{Relic density}
\label{sec:relic}

\begin{figure}[t!]
\begin{center}
\includegraphics[width=0.625\textwidth]{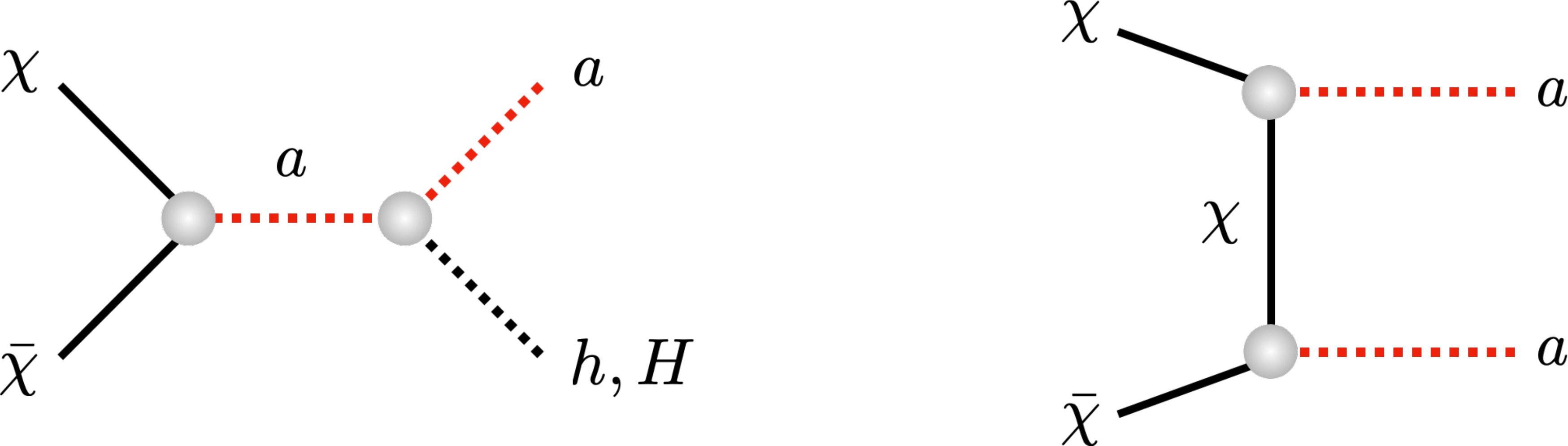}
\end{center}
\vspace{0mm} 
\caption{\label{fig:diagrams2} Feynman diagrams that lead to DM annihilation via $\chi \bar \chi \to ah$ or $\chi \bar \chi \to aH$~(left) and $\chi \bar \chi \to aa$ in the 2HDM+$a$~model. The possible decay modes of the pseudoscalar $a$, the SM-like Higgs $h$ and the heavy CP-even Higgs $H$ are not shown. Further details are given in the main text.}
\end{figure}

In order to understand the physics of standard thermal relic freeze-out in 2HDM+$a$ realisations with $\sin \theta \simeq 0$, we first write the cross section for annihilation of DM into a~final state~$X$~as
\beq \label{eq:TavgXS}
\sigma \left ( \chi \bar \chi \to X \right ) v_{\rm rel} = \sigma^0_X + \sigma^1_X \hspace{0.5mm} v_{\rm rel} ^2 \,,
\eeq
where $v_{\rm rel}$ is the relative velocity of the DM pair and the coefficient~$\sigma^0_X$ ($\sigma^1_X$) describes the $s\hspace{0.25mm}$-wave ($p\hspace{0.25mm}$-wave) contribution. 

In the alignment limit, the possible DM annihilation channels involving a pseudoscalar~$a$ are $\chi \bar \chi \to a \to f \bar f$, $\chi \bar \chi \to a \to ZH$, $\chi \bar \chi \to a \to ah$ and $\chi \bar \chi \to a \to aH$ for what concerns $s$-channel processes and $\chi \bar \chi \to a a$ with DM exchange in the $t$-channel (cf.~also \cite{LHCDarkMatterWorkingGroup:2018ufk}). The~annihilation cross sections~(\ref{eq:TavgXS}) of the former two reactions are, however, proportional to $\sin^2 \theta$ making them numerically irrelevant in the limit $\sin \theta \to 0$ unless $m_a = m_\chi/2$. Such~highly tuned solutions to the DM miracle will not be considered in what follows. Similarly, all DM annihilation contributions involving the exchange of a heavy pseudoscalar~$A$ are suppressed by at least two powers of the sine of the mixing angle $\theta$, so that only the processes depicted in~Figure~\ref{fig:diagrams2} are relevant for the calculation of the DM abundance in the context of this~work. 

The annihilation process $\chi \bar \chi \to a \to ah$ proceeds via $s$-wave and we find for the corresponding coefficient the following analytic result
\beq \label{eq:swaveah}
\sigma^0_{ah} = \frac{y_\chi^2 \hspace{0.5mm} g_{haa}^2 \cos^2 \theta}{32\hspace{0.125mm} \pi} \, \sqrt{1 - \frac{\left (m_h + m_a \right ) ^2}{4 m_\chi^2} } \, \sqrt{1 - \frac{\left (m_h - m_a \right ) ^2}{4 m_\chi^2} } \, \frac{v^2}{\left (m_a^2 - 4 m_\chi^2 \right )^2 + m_a^2 \hspace{0.25mm} \Gamma_a^2} \,,
\eeq
where the expression for $g_{haa}$ in the limit $\sin \theta \to 0$ can be found in the first line of~(\ref{eq:ghHaa}) and~$\Gamma_a$ denotes the total decay width of the pseudoscalar $a$. Since $\sigma^0_{ah} \neq 0$ we ignore the $p\hspace{0.25mm}$-wave coefficient $\sigma^1_{ah}$ below by setting it to zero. The result for the $s\hspace{0.25mm}$-wave coefficient~$\sigma^0_{aH}$ describing DM annihilation through $\chi \bar \chi \to a \to aH$ is simply obtained from~(\ref{eq:swaveah}) by the replacements $g_{haa} \to g_{Haa}$ and $m_h \to m_H$. 

In the case of $\chi \bar \chi \to aa$ the annihilation cross section is instead $p$-wave suppressed (see~\cite{Albert:2017onk} for the calculation of the $t$-channel contribution in the simplified pseudoscalar DM model) and the corresponding expansion coefficients take the form $\sigma^0_{aa} = 0$ and 
\beq \label{eq:pwaveaa}
\sigma^1_{aa} = \frac{y_\chi^4 \hspace{0.25mm} \cos^4 \theta}{24 \hspace{0.125mm} \pi} \, \sqrt{1 - \frac{m_a^2}{m_\chi^2} } \, \frac{m_\chi^2 \left (m_a^2 - m_\chi^2 \right )^2}{\left (m_a^2 - 2 m_\chi^2 \right )^4} \,.
\eeq

\begin{figure}[t!]
\begin{center}
\includegraphics[width=0.6\textwidth]{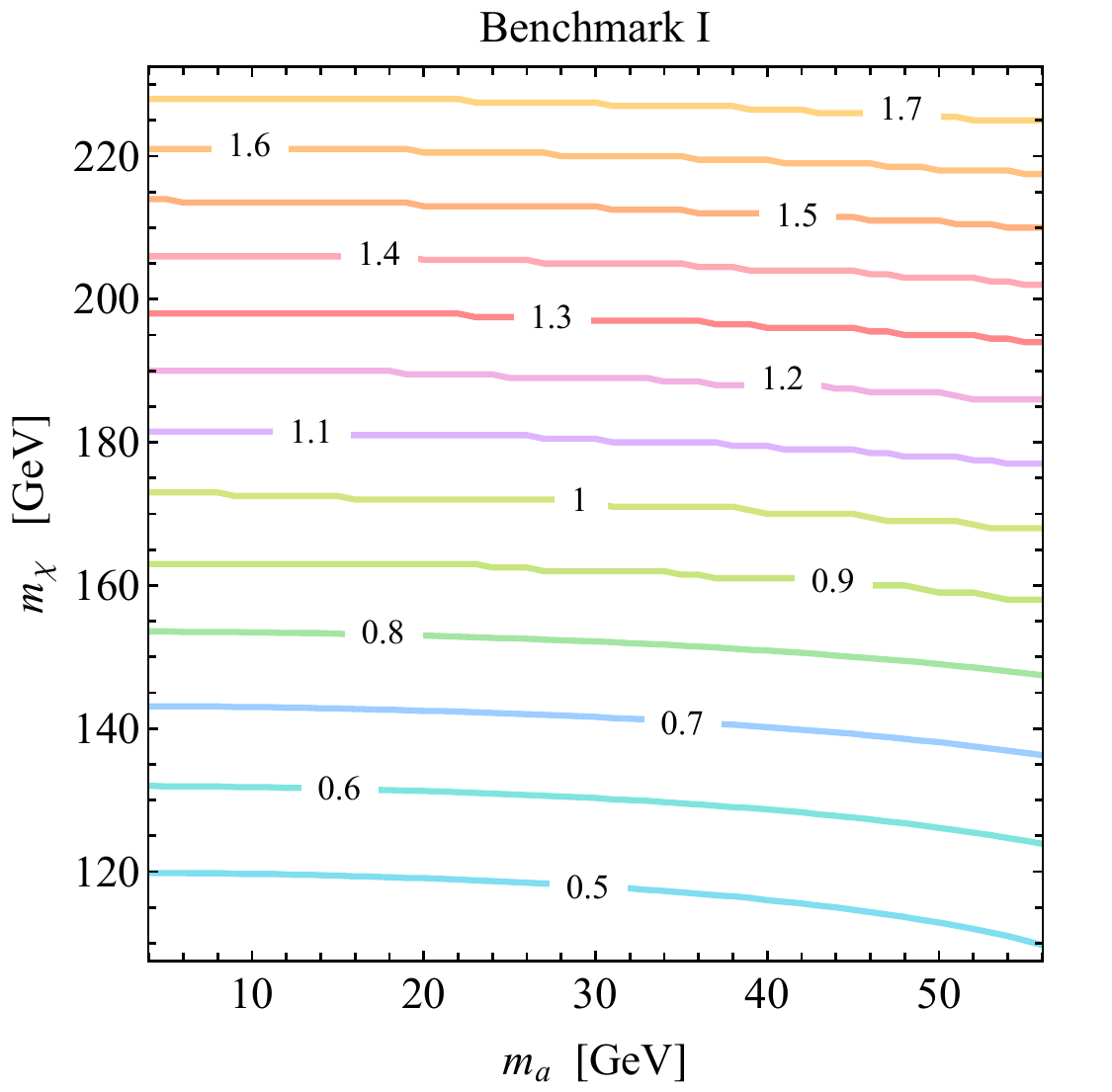}
\end{center}
\vspace{-4mm} 
\caption{\label{fig:summary4} Predicted DM relic abundance in the $m_a\hspace{0.125mm}$--$\hspace{0.5mm}m_\chi$ plane for the 2HDM+$a$ benchmark~I~parameter choices~(\ref{eq:benchmarkI}). The contour lines indicate the value of~$\Omega h^2/0.12$, meaning that the regions below (above) 1 correspond to a DM underabundance (overabundance) in today's Universe. For~additional details we refer the interested reader to the main text.}
\end{figure}

Using the velocity expansion~(\ref{eq:TavgXS}) the DM relic density after freeze-out can be approximated by 
\beq \label{eq:Omegah2}
\frac{ \Omega h^2}{0.12} \simeq \frac{1.6 \cdot 10^{-10} \, {\rm GeV}^{-2} \, x_f}{\left \langle \sigma \hspace{0.125mm} v_{\rm rel} \right \rangle_f} \,, \qquad 
\left \langle \sigma v_{\rm rel} \right \rangle_f = \sum_X \left ( \sigma_X^0 + \frac{3 \hspace{0.25mm} \sigma_X^1}{x_f} \right ) \,.
\eeq
Here $x_f = m_\chi/T_f \in [20, 30]$ with $T_f$ the freeze-out temperature and the sum over $X$ in principle includes all possible final states. As we have explained above, for $\sin \theta \simeq 0$ and away from the exceptional points $m_a=m_\chi/2$, however, only the channels $X = ah, aH, aa$ are numerically important. In the limit of heavy DM,~i.e.~$m_\chi \gg m_a, m_h, m_H$, the velocity-averaged annihilation cross section at the freeze-out temperature can be further simplified:
\beq \label{eq:Omegah2denapp}
\left \langle \sigma v_{\rm rel} \right \rangle_f \simeq \frac{y_\chi^2}{128\hspace{0.25mm} \pi \hspace{0.25mm} m_\chi^2} \left [ \frac{\left ( g_{haa}^2 + g_{Haa}^2 \right ) \hspace{0.125mm} v^2}{4 \hspace{0.25mm} m_\chi^2} + \frac{y_\chi^2}{x_f} \right ] \,. 
\eeq
This approximation shows that the $s$-channel ($t$-channel) contributions to $\left \langle \sigma v_{\rm rel} \right \rangle_f$ scale as $1/m_\chi^4$~$\left (1/m_\chi^2 \right )$ in the limit of infinitely heavy DM. 

\begin{figure}[t!]
\begin{center}
\includegraphics[width=0.6\textwidth]{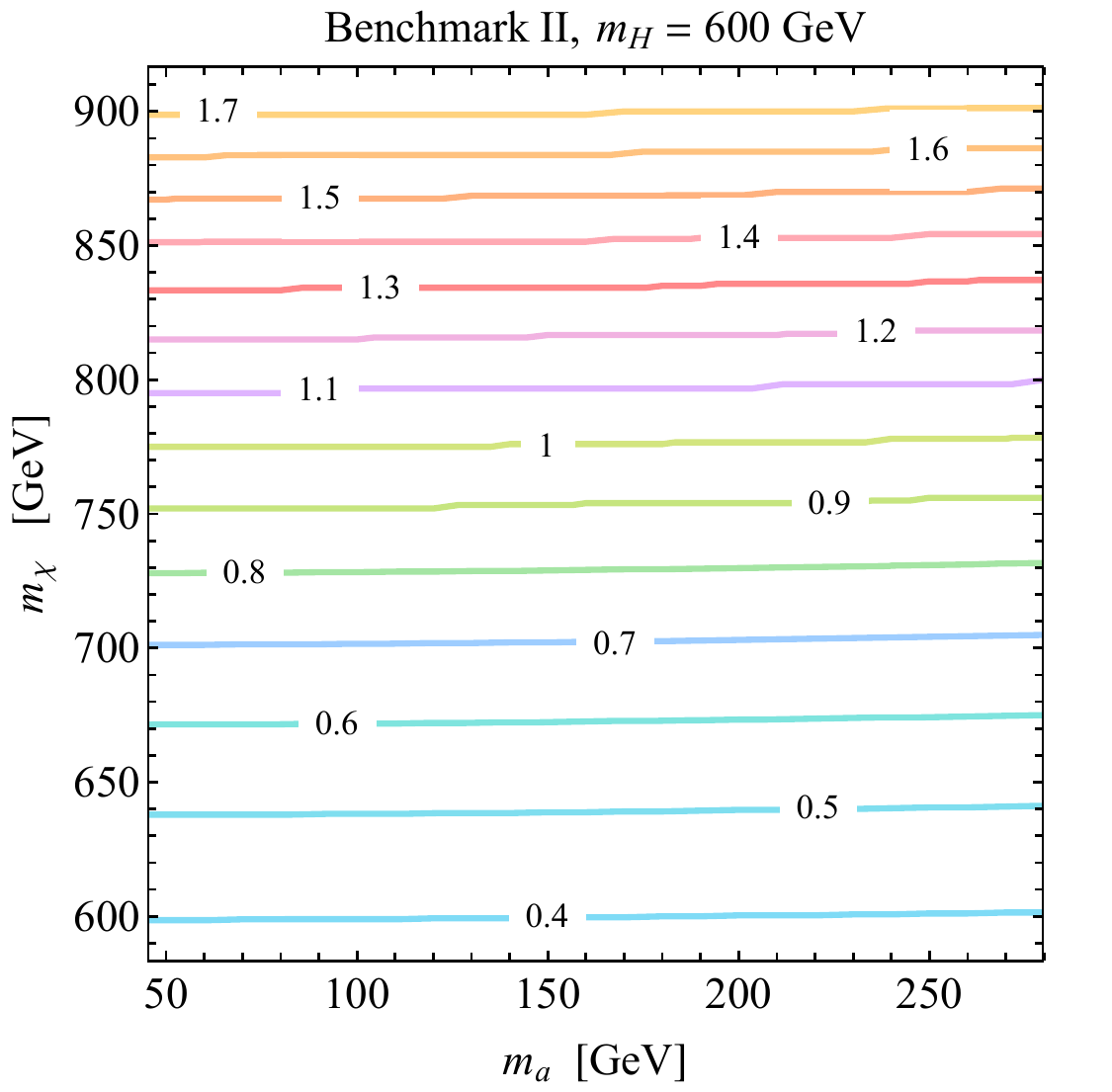}

\hspace{6mm}

\includegraphics[width=0.6\textwidth]{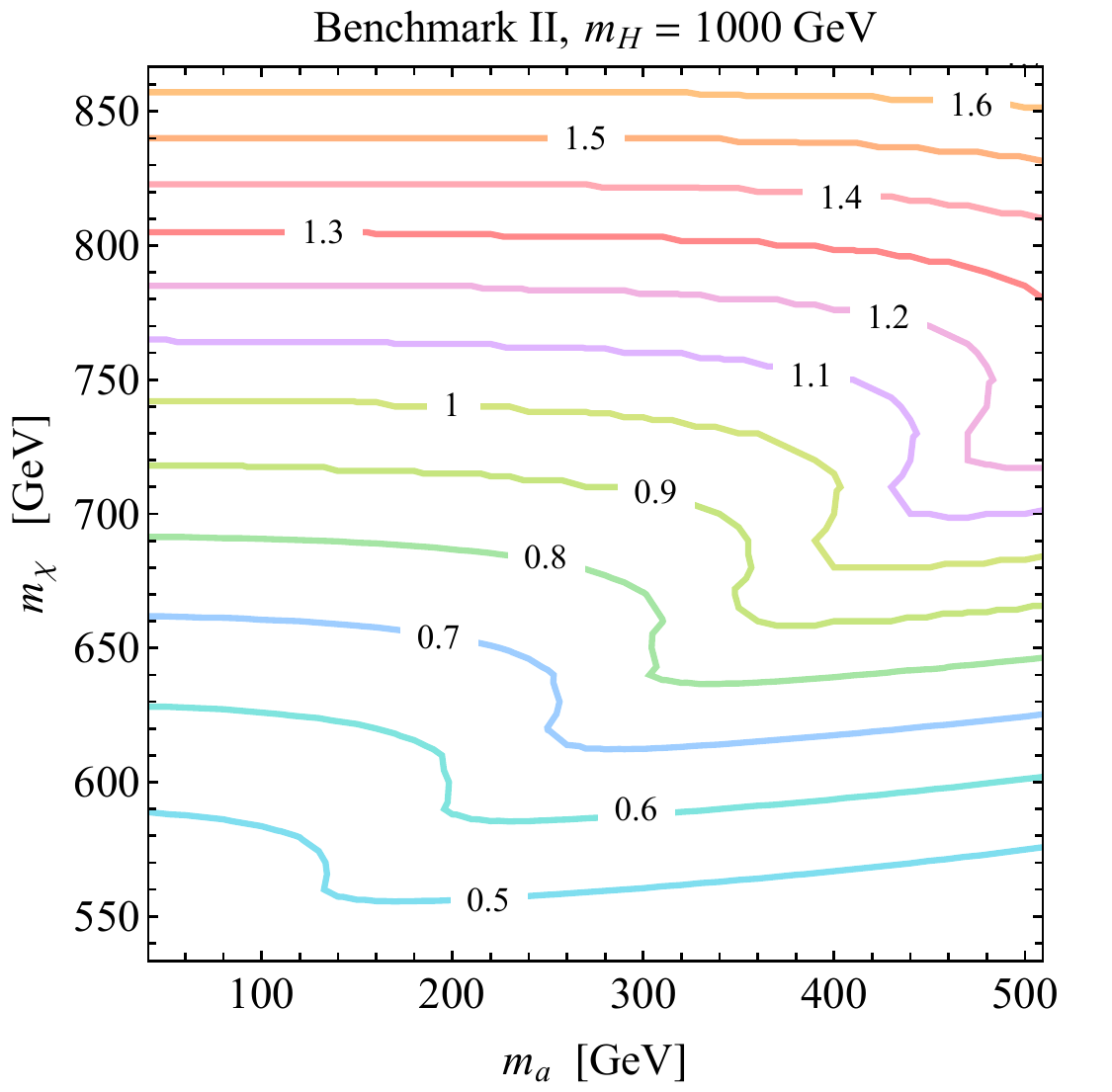}
\end{center}
\vspace{-4mm} 
\caption{\label{fig:summary56} As~Figure~\ref{fig:summary4} but for the 2HDM+$a$ benchmark~II scenario with $m_H = 600 \, {\rm GeV}$~(upper panel) and $m_H=1000 \, {\rm GeV}$~(lower panel), respectively. Additional details can again be found in the main text.}
\end{figure}

The formulas~(\ref{eq:Omegah2}) and~(\ref{eq:Omegah2denapp}) represent useful expressions to estimate~$\Omega h^2$. In the case of the benchmark~I scenario~(\ref{eq:benchmarkI}) one has $g_{haa}^2 \simeq 6 \cdot 10^{-5}$ and $g_{Haa}^2 = 0$, and it is thus a good approximation to neglect the $s$-channel contributions to $\left \langle \sigma v_{\rm rel} \right \rangle_f$. It follows that 
\beq \label{eq:Omegah2benchIapp}
\frac{ \Omega h^2}{0.12} \simeq \frac{0.9}{y_\chi^4} \left ( \frac{x_f}{25} \right )^2 \left ( \frac{m_\chi}{150 \, {\rm GeV}} \right )^2 \,.
\eeq
Using $x_f \simeq 25$ the relic abundance of~$\Omega h^2 = 0.120 \pm 0.001$ as determined by Planck~\cite{Planck:2018vyg} is therefore obtained in the case of~(\ref{eq:benchmarkI}) for DM masses $m_\chi \simeq 160 \, {\rm GeV}$ while for parameter regions with $m_\chi \lesssim 160 \, {\rm GeV}$ ($m_\chi \gtrsim 160 \, {\rm GeV}$) one expects DM underabundance (overabundance). These expectations agree quite well with the results of our exact relic calculation that have been obtained with {\tt MadDM}~\cite{Backovic:2015tpt} and are shown in~Figure~\ref{fig:summary4}. In~fact, the exact computation for~(\ref{eq:benchmarkI}) and $m_a = 30 \, {\rm GeV}$ leads to~$\Omega h^2 = 0.118$, while~(\ref{eq:Omegah2benchIapp}) naively predicts a value that is larger by around 15\%. The observed difference can be traced back to the fact that the {\tt MadDM} calculation gives $x_f \simeq 21$ in the parameter region of interest and correctly takes into account the phase-space suppression present in~(\ref{eq:pwaveaa}) due to the non-zero values of $m_a^2/m_\chi^2$. We add that $\chi \bar \chi \to a \to a h$ annihilation represents a relative contribution to~$\Omega h^2$ of less than about 1\% in the part of the $m_a\hspace{0.125mm}$--$\hspace{0.5mm}m_\chi$ plane that is depicted in the~figure. Neglecting the $s$-channel contributions in the approximation~(\ref{eq:Omegah2benchIapp}) is hence fully~justified. 

In the case of our 2HDM+$a$ benchmark~II parameter scenario~(\ref{eq:benchmarkII}) the coupling $g_{haa}$ is no longer small, in fact $g_{haa}^2 \simeq 35$ and we furthermore have $g_{Haa}^2 \neq 0$. On the other hand, $y_\chi^2/x_f \simeq 0.04$ and thus one can neglect the $\chi \bar \chi \to a a$ contribution to the velocity-averaged annihilation cross section at the freeze-out temperature~(\ref{eq:Omegah2denapp}) to first approximation. Doing this, one obtains the following simple expression 
\beq \label{eq:Omegah2benchIIapp}
\frac{ \Omega h^2}{0.12} \simeq \frac{43}{y_\chi^2 \left ( g_{haa}^2 + g_{Haa}^2 \right ) } \, \frac{x_f}{25} \left ( \frac{m_\chi}{800 \, {\rm GeV}} \right )^4 \,,
\eeq
which approximately describes the resulting DM relic abundance for parameter choices {\`a}~la benchmark~II. From~(\ref{eq:Omegah2benchIIapp}) one hence expects that the correct value of~$\Omega h^2$ is realised in the case of~(\ref{eq:benchmarkII}) for $m_\chi \simeq 770 \, {\rm GeV}$, while for smaller (larger) DM masses one should have~$\Omega h^2 \gtrsim 0.12$~$\left ( \Omega h^2 \lesssim 0.12 \right)$ if $m_H = 600 \, {\rm GeV}$ which implies $g_{Haa}^2\simeq 1.5$. The~results of the corresponding {\tt MadDM} computation is displayed in the upper panel of~Figure~\ref{fig:summary56}. We~find that for the parameters~(\ref{eq:benchmarkII}) together with $m_a = 150 \, {\rm GeV}$ as well as $m_H = 600 \, {\rm GeV}$, the exact calculation predicts~$\Omega h^2 = 0.117$, a value less than 5\% below the naive expectation. Numerically, we furthermore obtain that the relative contribution of $\chi \bar \chi \to a a$ to the DM relic density is always below $1\%$ in benchmark~II with $m_H = 600 \, {\rm GeV}$, showing that it can be safely neglected in the derivation of~(\ref{eq:Omegah2benchIIapp}). 

The result of our ${\tt MadDM}$ scan in the 2HDM+$a$ benchmark~II parameter scenario with $m_H = 1000 \, {\rm GeV}$ is presented in the lower panel of~Figure~\ref{fig:summary56}. It is evident from the~plot that in this case the above simplistic formula is not able to capture the more intricate behaviour of the contours of constant relic density. In fact, this is not a big surprise because in the derivation of~(\ref{eq:Omegah2benchIIapp}) we have assumed that $m_\chi \gg m_a, m_h, m_H$, however, one has $m_\chi < m_H$ in the entire $m_a\hspace{0.125mm}$--$\hspace{0.5mm}m_\chi$ plane considered. Still the values $m_\chi \in [680, 740] \, {\rm GeV}$ of the DM mass that lead to the correct DM abundance for $m_a \in [50, 500] \, {\rm GeV}$ are only less than 10\% smaller than what one would expect from~(\ref{eq:Omegah2benchIIapp}). Like in the case before, it turns out that the relative contribution of $s$-channel annihilation amounts to more than~99\% of the predicted values of~$\Omega h^2$. This shows again that DM annihilation via $\chi \bar \chi \to aa$ is phenomenologically irrelevant in 2HDM+$a$ realisations of the type~(\ref{eq:benchmarkII}). 

Before concluding, we add that DM direct detection experiments do not set relevant constraints on the 2HDM+$a$ benchmarks~(\ref{eq:benchmarkI}) and (\ref{eq:benchmarkII}) for the mixing angles $\theta \simeq 0$ necessary to have a long-lived $a$. This is a simple consequence of the fact that the spin-independent DM-nucleon cross section is suppressed by both a loop factor and two powers of $\sin \theta$ --- see the recent articles~\cite{Argyropoulos:2022ezr,Arcadi:2017wqi,Bell:2018zra,Abe:2018emu,Ertas:2019dew} for explicit formulas and further explanations. 

\section{Conclusions}
\label{sec:summary}

The main lesson that can be learnt from the analytic and numerical results presented in this work is that LHC searches for displaced Higgs decays can provide unique constraints on 2HDM+$a$ model realisations. In fact, LLP signatures appear in the context of the 2HDM+$a$ model if the mixing angle $\theta$ of the two CP-odd weak spin-0 eigenstates is very small and the DM sector is either decoupled or kinematically inaccessible. In order to emphasise this generic finding, we have studied two distinct parameter benchmarks and explored the sensitivity of the existing LHC LLP searches by performing parameter scans in the $m_a\hspace{0.25mm}$--$\hspace{0.5mm} \sin \theta$~plane. The results of these scans can be found in~Figures~\ref{fig:summary1} and~\ref{fig:summary23}. 

In~the~benchmark~I scenario we have chosen the 2HDM+$a$ parameters such that the $125 \, {\rm GeV}$ Higgs boson gives rise to an LLP signature through its prompt decay $h \to aa$ followed by the displaced decays of the pseudoscalars to SM fermions such as $a \to b \bar b$ or gluons. One~important feature that is nicely illustrated in our benchmark~I scan is that the LLP searches for displaced hadronic jets that have been performed at LHC~Run~II can probe regions of parameter space with mixing angles $\theta$ in the ballpark of $10^{-7}$ to $10^{-5}$ that are presently not accessible by any other means. In fact, in our benchmark~I scenario the predicted $h \to aa$ branching ratio turns out to be below the target sensitivity that the HL-LHC is expected to reach on undetected or invisible decays of the $125 \, {\rm GeV}$ Higgs boson. 2HDM+$a$ realisations like~(\ref{eq:benchmarkI}) are therefore unlikely to be testable indirectly at the LHC in searches for both prompt $h \to aa \to 4f$ production or through signatures involving a significant amount of $E_{T, \rm miss}$ such as $h \to {\rm invisible}$ or mono-jet final states. 

The benchmark~II parameters were instead chosen such that the LLP signal arises from the prompt decay $H \to aa$ of the heavy CP-even Higgs followed by the displaced decays of the pseudoscalars to a pair of SM fermions or gluons. Depending on the precise value of the LLP mass either the $a \to b \bar b$ or the $a \to t \bar t$ mode turns out to provide the largest rate. In order to illustrate these two distinct possibilities, we have considered in benchmark~II the mass combination $m_H = 600 \, {\rm GeV}$ with $m_a \in [50, 275] \, {\rm GeV}$ as well as $m_H = 1000 \, {\rm GeV}$ with $m_a \in[50, 475] \, {\rm GeV}$. In the former case we found that the existing LHC searches for displaced heavy Higgs decays provide stringent constraints on $\theta$, excluding values of the mixing angle between $2 \cdot 10^{-8}$ and $2 \cdot 10^{-6}$. The limits on the benchmark~II scenario with $m_H = 1000 \, {\rm GeV}$ 
turned out to be noticeably weaker than the bounds for $m_H = 600 \, {\rm GeV}$ due to the order of magnitude smaller inclusive heavy Higgs production cross section. Still, for~the three mass values $m_a = 50 \, {\rm GeV}, 150 \, {\rm GeV}$ and $275 \, {\rm GeV}$, $\theta$ values between~$4 \cdot 10^{-8}$ and~$2 \cdot 10^{-6}$ are excluded, whereas for $m_a = 400 \, {\rm GeV}$ mixing parameters in the range of~$4 \cdot 10^{-9}$ and~$1 \cdot 10^{-9}$ are disfavoured by the searches for the displaced heavy Higgs decays performed at LHC~Run~II. We expect that LLP searches at LHC~Run~III and beyond will provide sensitivity to 2HDM+$a$ models {\`a}~la~(\ref{eq:benchmarkII}) with heavy CP-even Higgs masses of the order of~$1 \, {\rm TeV}$ and pseudoscalar masses $m_a$ above the top-quark threshold. Let us finally note that in future LHC runs it should also be possible to probe 2HDM+$a$ benchmark~II models with a heavy CP-even Higgs that satisfies $m_H > 2 m_a$ and $m_H > 2 m_t$ through spin-0 resonance searches in di-top and four-top production. A detailed analysis of this issue is however beyond the scope of this work. 

We have furthermore demonstrated that parameter choices that give rise to an interesting LLP phenomenology can simultaneously explain the observed DM abundance without excessive tuning. Our relic density scans have been given in~Figures~\ref{fig:summary4} and~\ref{fig:summary56}. In our benchmark~I scenario, we have seen that due to the smallness of the $g_{haa}$ coupling the value of~$\Omega h^2$ is entirely determined by the $t$-channel process $\chi \bar \chi \to aa$ which is $p\hspace{0.25mm}$-wave suppressed. For~the range $m_a < m_h/2$ relevant in benchmark~I, we found that the correct relic abundance can be obtained for DM masses of around $170 \, {\rm GeV}$, rather independently of the precise value of $m_a$. In the case of the benchmark~II model instead, only the $s$-channel annihilation contributions $\chi \bar \chi \to a \to ah$ and $\chi \bar \chi \to a \to aH$ that are $s\hspace{0.25mm}$-wave turn out to be phenomenologically relevant. While the precise value of the DM mass for which~$\Omega h^2 \simeq 0.12$ is achieved depends on the chosen parameters such as $m_a$ and~$m_H$, we saw that DM masses in the range $m_\chi \in [680, 770] \, {\rm GeV}$ allow to saturate the measured relic abundance in the part of the parameter space most relevant for the LLP signals triggered by $H \to aa$. 

The results and parameter benchmarks presented in our article should provide a useful starting point for interpretations of future ATLAS, CMS and LHCb searches for displaced Higgs decays to hadronic jets in the context of the 2HDM+$a$ model. We therefore encourage and look forward to experimental explorations in this direction. 

\acknowledgments UH thanks Spyros Argyropoulos for interesting discussions that triggered this research. His~comments on the manuscript and his questions concerning the DM phenomenology of the model where also very valuable in improving our work. The research of~LS is partially supported by the International Max Planck Research School~(IMPRS) on ``Elementary Particle Physics'' as well as the Collaborative Research Center SFB1258.



%

\end{document}